\documentclass[12pt]{article}
\usepackage{amsmath,amssymb,amsthm,amsfonts}
\usepackage{mathtools}
\usepackage{graphicx}
\usepackage{enumerate}
\usepackage{natbib}
\usepackage{url} 
\usepackage[justification=centering]{caption}
\usepackage{hyperref}

\usepackage{algpseudocode}
\usepackage{algorithm}

\newtheorem{mydef}{Definition}
\newtheorem{theorem}{Theorem}
\newtheorem{lemma}{Lemma}

\newcommand\mydistr{\stackrel{\mathclap{\normalfont\mbox{\small{d}}}}{=}}
\newcommand\med{\text{med}}

\usepackage{tikz}
\usetikzlibrary{arrows,positioning}
\usetikzlibrary {arrows.meta}
\usepackage{tkz-graph}
\tikzset{
  LabelStyle/.style = { rectangle, rounded corners, draw,
                        minimum width = 2em, fill = yellow!50,
                        text = red, font = \bfseries },
  VertexStyle/.append style = { minimum size = 35pt,inner sep=3pt,                                font = \normalsize\bfseries, color = black, fill=white,line width=0.5mm},
  EdgeStyle/.append style = {->,line width=0.5mm} 
  }
  \usetikzlibrary{arrows.meta}

\begin{document}

\def\spacingset#1{\renewcommand{\baselinestretch}%
{#1}\small\normalsize} \spacingset{1}

  \title{\bf TSLiNGAM: DirectLiNGAM under heavy tails}
  \author{Sarah Leyder\thanks{
    First author is supported by Fonds Wetenschappelijk onderzoek - Vlaanderen (FWO) as a
    PhD fellow Fundamental Research (PhD fellowship 11K5523N).}\hspace{.2cm}\\
    Department of Mathematics, University of Antwerp,\\
    Jakob Raymaekers \\
    Department of Quantitative Economics, Maastricht University, \\
    Tim Verdonck \\
    Department of Mathematics, University of Antwerp - imec\\
    Department of Mathematics, KU Leuven}
    
  \maketitle

\bigskip
\begin{abstract}
One of the established approaches to causal discovery consists of combining directed acyclic graphs (DAGs) with structural causal models (SCMs) to describe the functional dependencies of effects on their causes. Possible identifiability of SCMs given data depends on assumptions made on the noise variables and the functional classes in the SCM. For instance, in the LiNGAM model, the functional class is restricted to linear functions and the disturbances have to be non-Gaussian. 

In this work, we propose TSLiNGAM, a new method for identifying the DAG of a causal model based on observational data. TSLiNGAM builds on DirectLiNGAM, a popular algorithm which uses simple OLS regression for identifying causal directions between variables. TSLiNGAM leverages the non-Gaussianity assumption of the error terms in the LiNGAM model to obtain more efficient and robust estimation of the causal structure.
TSLiNGAM is justified theoretically and is studied empirically in an extensive simulation study. It performs significantly better on heavy-tailed and skewed data and demonstrates a high small-sample efficiency. In addition, TSLiNGAM also shows better robustness properties as it is more resilient to contamination.
\end{abstract}

\noindent%
{\it Keywords:}  Causal discovery, Efficiency, LiNGAM, Structural causal models 
\vfill

\newpage
\spacingset{1.5} 
\section{Introduction}
\label{sec:intro}

Over the last decades, statistics and machine learning have proven to be very strong tools for finding and modeling associations in data sets. However, in recent years, it has become clear that when analyzing and using data, causal relations are often more valuable than associations. As a result, there has been a growing interest in causal inference in statistics and machine learning, and it has become a crucial tool in many empirical sciences including medicine, social sciences, neuroinformatics and biology. 

One of the established approaches to causal inference builds on the directed acyclic graph (DAG) framework, studied in great depth by \cite{Pearl}. DAGs represent the variables of interest as nodes in a graph where directed edges between nodes correspond with causal relations. As DAGs are acyclic, i.e. there are no cycles in the network, they represent a one-directional causal order of the variables, such that no variable with a later causal order can influence an earlier variable. DAGS are typically complemented by structural causal models (SCMs), which are used to describe the functional dependence of an effect on its causes in a DAG. This combined theoretical framework then bestows us with the necessary tools to compute observational, interventional and counterfactual distributions for answering all possible causal queries.

SCMs describe the causal relationships among random variables $X_1,\dots,X_p$ as a set of equations
\begin{equation*}
    X_i = f_i(\text{P}_i,e_i), \qquad i = 1, \dots, p
\end{equation*}
where the variable set $\text{P}_i \subset \{ X_1, \dots, X_p \}$ denotes the parent variables of $X_i$, i.e. the variables that determine $X_i$, and $e_i$ represents the random noise variable disturbing $X_i$. A central question in causal inference is whether the SCM can be identified from data. In other words, given enough data, can we find the DAG (and the functional dependencies) underlying the data generating process? It is known that this is impossible in full generality. However, under appropriate conditions on the behavior of the noise variables and the functional class of the $f_i$, it is indeed possible to recover the SCM from observational data alone \citep{peters2017elements}. This process is called causal discovery. It provides a very valuable addition to (randomized) controlled experiments, which are often difficult or impossible due to prohibitive costs or ethical objections.

One prime example of conditions under which identifiability is possible, is the LiNGAM model \citep{LiNGAM}. In LiNGAM, the functional class is restricted to linear functions and the disturbances have to be non-Gaussian and mutually independent. Under these assumptions, identifiability is provable and given continuous data, the complete causal structure can be recovered. The above described model is known as the linear, non-Gaussian, acyclic model (LiNGAM) and the original discovery algorithm is based on independent component analysis (ICA), justified by the assumption of non-Gaussianity \citep{LiNGAM}. Several extension to the LiNGAM model have been made. LvLiNGAM \citep{lvLiNGAM} includes the presence of hidden variables or latent confounders by using overcomplete ICA, other methods resilient against latent confounders are ParceLiNGAM \citep{parceLiNGAM} and MLCLiNGAM \citep{MLCLiNGAM}. \cite{LiNGAMmeasurementerror} consider causal discovery of the LiNGAM model in the presence of measurement errors. \cite{timeseries} discuss the integration of LiNGAM in autoregressive models for time series.

In this work, we propose TSLiNGAM, a new algorithm for estimating the causal structure in a LiNGAM model. TSLiNGAM builds on DirectLiNGAM \citep{DirectLiNGAM}, which is a popular method to obtain the causal LiNGAM model based on simple OLS regressions, but relies on regression estimators which are more efficient under heavy tails and skewness. These alternative regression estimators are more natural given the non-gaussianity assumption in the LiNGAM model, and their appropriateness is further motivated theoretically and empirically. 

The remainder of the article is organized as follows. Section \ref{sec:method} briefly reviews DirectLiNGAM and introduces TSLiNGAM. It also discusses the theoretical and computational properties of TSLiNGAM. Section \ref{sec:simulation} then demonstrates the advantage of TSLiNGAM over DirectLiNGAM in an extensive simulation study. Lastly, in Section \ref{sec:realdata}, we compare TSLiNGAM to DirectLiNGAM on four real data examples. Finally, Section \ref{sec:conclusion} concludes.

\section{Method}
\label{sec:method}
We start by reviewing the LiNGAM model and the DirectLiNGAM algorithm, before introducing TSLiNGAM.

\subsection{Preliminaries}
The LiNGAM structural causal model postulates that the functional dependencies are linear and the external influences are independent and non-Gaussian. More precisely, it relies on the following three assumptions:
\begin{enumerate}
    \item The generating process can be described by a directed acyclic graph such that the variables $\{X_1, \dots, X_p\}$ can be arranged in a causal order. The causal order of the variable $X_i$ is denoted by $k(i)$.
    \item Each variable is a linear combination of other variables with a lower causal order, plus an external influence:
    \begin{equation*}
        X_i = \sum_{k(j) < k(i)} b_{ij} X_j + e_i
    \end{equation*}
    The coefficients $b_{ij}$, called the connection strengths, can be arranged into a matrix $B$, which can be permuted to strict lower triangularity since the generating process concerns a DAG. The noise terms $e_i$ can be placed into a vector $e$. Hence we obtain the matrix notation:
    \begin{equation}
    \label{eq:LiNGAM}
        X = BX + e
    \end{equation}
    We call $X_i$ an exogenous variable if $X_i$ is equal to $e_i$, so no variable $X_j$ has a directed path to $X_i$. In the DAG framework, there is always at least one exogenous variable. Non-exogenous variables are called endogenous variables.
    \item The external influences $e_i$ are continuous random variables following a non-Gaussian distribution with zero mean and non-zero variance and all the $e_i$ for $i \in \{1,\dots,p\}$ are independent of each other.
\end{enumerate}

Given a data set, the underlying LiNGAM structure can be recovered by rewriting Equation \eqref{eq:LiNGAM} as:
\begin{equation}
\label{ICAeq}
    X = Ae
\end{equation}
where $A = (I-B)^{-1}$. Since the disturbance vector $e$ contains mutually independent, non-Gaussian variables, Equation \eqref{ICAeq} corresponds to the well-known linear independent component analysis model (ICA). The matrix $A$ is called the mixing matrix and efficient ICA-algorithms exist to estimate it for a given data set. Subsequently scaling and permutation steps can be performed to produce a strictly lower triangular matrix $B$, from which the corresponding causal order can then easily be derived. More information on the LiNGAM discovery algorithm can be found in the original LiNGAM paper \citep{LiNGAM}.

The LiNGAM algorithm using ICA-estimation does, however, have some drawbacks. First, the optimization used for ICA can get trapped in a local minimum and hence we have no guaranteed computational stability for the method. Second, for the gradient-based algorithm, appropriate parameters must be selected which is not easily done.

In 2011, a direct method was proposed to estimate causal ordering in the linear non-Gaussian context, namely DirectLiNGAM \citep{DirectLiNGAM}. In contrast to ICA-LiNGAM, this new method has guaranteed convergence and requires no parameter specification. DirectLINGAM uses two main ingredients. The first is OLS regression to remove the effect of an exogenous variable from the other variables. The second is an independence measure to identify the next exogenous variable.  Denote with $r_i^{(j)} \coloneqq X_i - \frac{\text{cov}(X_i,X_j)}{\text{var}(X_j)} X_j$ the ordinary least squares residual when $X_i$ is regressed on $X_j$. Further denote the kernel-based estimator of mutual information \citep{BachJordan} with $\widehat{MI}_{\text{kernel}}$.  For each variable $X_j$ we sum the mutual information of it with each of its ordinary least squares residuals $r_i^{(j)}$ to obtain the kernel-based independence measure (KBI):
\begin{equation}
\label{eq:KBIM}
    T_{\text{kernel}}(X_j,U) = \sum_{i \in U, \ i \neq j} \widehat{MI}_{\text{kernel}}(X_j, r_i^{(j)})
\end{equation}
Here $U$ is the set of indices of the remaining variables. The variable with the lowest $T_{\text{kernel}}$ is then the most independent and will be used as the next exogenous variable.

In summary, the DirectLiNGAM algorithm proceeds as follows:\\
\begin{algorithm}[hbt!]
\caption{DirectLiNGAM algorithm \citep{DirectLiNGAM}}\label{alg:DirectLiNGAM}
\begin{algorithmic}
  \scriptsize
\Require $n \times p$ data set $\mathbf{X}$
\State $U \gets \{1,\dots,p\}$  \Comment{Initialize the set of variable subscripts}
\State $K \gets \emptyset$ \Comment{Initialize an empty ordered list of variable subscripts}

\While{$K$ contains less than $p-1$ indices}
\For{$j \in U\setminus K$}\Comment{Cycle through the variables in $U\setminus K$}
\For{$i \in U\setminus (K \cup j)$}\Comment{Cycle through the variables in $U\setminus (K\cup j)$}
\State $R_{\cdot,i}^{(j)} \gets r_i^{(j)}$ \Comment{Store the OLS residuals of variable $X_i$ on $X_j$}
\EndFor
\State $T_j \gets T_{\text{kernel}}(X_j,U \setminus K)$
\EndFor
\State $m = \text{argmin}_{j \in U \setminus K} \ T_j$ \Comment{Find the next exogenous variable}
\State $K \gets \{K, m\}$ \Comment{Append $m$ to $K$}
\State $X \gets r^{(m)}, \mathbf{X} \gets R^{(m)}$ \Comment{Consider the residuals as new input}
\EndWhile
\State $K \gets \{K, (U\setminus K)\}$ \Comment{Append the final variable to obtain the complete causal ordering}
\State $B \gets \mbox{OLS}(\mathbf{X},K)$ \Comment{perform OLS on $\mathbf{X}$ following the order in $K$}
\end{algorithmic}
\end{algorithm}

\noindent
The algorithm above can be extended to make use of prior knowledge on the structure if this is available. For more details, we refer to the DirectLiNGAM paper \citep{DirectLiNGAM}. As is clear from the pseudo-code description in Algorithm \ref{alg:DirectLiNGAM}, DirectLiNGAM relies crucially on least squares regression. In addition, the proofs for the identification of the LiNGAM structure by DirectLiNGAM also rely on the use of the least squares estimator \citep{DirectLiNGAM}.\\
The DirectLiNGAM algorithm as introduced so far only identifies the causal ordering and returns a fully connected DAG, which is the focus of this paper. In order to drop redundant edges, it can be followed by a sparse regression estimator, for which the adaptive lasso \citep{zou2006adaptive} was used by \cite{DirectLiNGAM}. 

\subsection{TSLiNGAM}
\label{sec:TSLiNGAM}

The reliance of DirectLiNGAM on OLS regression is counterintuitive. OLS is known to perform extremely well under independent Gaussian errors, but loses its superiority when the errors are skewed, heavy tailed or heteroscedastic, especially when data samples are small \citep{Wilcox}. Given that the LiNGAM model assumes non-Gaussianity of the error terms, OLS is potentially a weak point of the algorithm.

In order to study this hypothesis, we propose the use of a different slope estimator to identify exogenous variables, namely the Theil-Sen regression estimator. This is motivated by its favorable properties on heavy-tailed and skewed distributions. Theil-Sen regression was first introduced by \cite{Theil} and later extended by \cite{Sen}. It is defined as follows:

\begin{mydef}[Theil-Sen slope]
For the linear regression of a random variable $Y$ on $X$, $Y = \beta X + e$, the Theil-Sen slope estimator is defined as
\begin{equation}
    \label{eq:TheilSen}
    \hat{\beta} = \text{med}_{i,j} \frac{y_j-y_i}{x_j - x_i}, \qquad \text{for } x_j \neq x_i
\end{equation}
for data pairs $\{ (x_i,y_i) : i = 1, \dots,n \}$.
\end{mydef}

The Theil-Sen slope estimator is unbiased, regression equivariant, robust with a breakdown value of 0.293 and a bounded influence function \citep{Sen, Peng}. Compared to OLS, it has a high small-sample efficiency and it is super-efficient when combined with discontinuous or discrete errors \citep{Wilcox, Peng}.  Also, when the errors are (close to) normal, Theil-Sen only loses little efficiency compared to OLS.

To apply the Theil-Sen slope in the DirectLiNGAM algorithm, we need to justify its use theoretically by generalizing the lemmas in \cite{DirectLiNGAM}. For this, we need
the functional form of the Theil-Sen given by
\begin{align}
\label{eq:theilsenfunctional}
    &T(X,Y) = \underset{X,X',Y,Y'}{\med} \left( \frac{Y - Y'}{X - X'} \right) = F^{-1}(0.5),
\end{align}
where $F$ denotes the distribution of $\frac{Y - Y'}{X - X'}$ with $Y \mydistr Y'$ and $X \mydistr X'$. Using this form, we first proof the Fisher Consistency of the Theil-Sen estimator, a property we will need later for generalizing the lemmas. It is defined as follows

\begin{mydef}[Fisher consistency]
The functional $T$ estimating a parameter $\Theta$ is Fisher consistent for a distribution $F$ if, when calculating the functional on the whole population, it equals the estimated parameter:
\begin{equation*}
    \text{for the distribution $F$: } T(F) = \Theta
\end{equation*}
\end{mydef}

It turns out that the Theil-Sen estimator is indeed Fisher consistent as shown in the result below

\begin{theorem}[Fisher consistency of Theil-Sen slope]
\label{thm:TSfishcons}
    For a simple linear regression model $Y = \beta X + \varepsilon$ such that $X$ and $\varepsilon$ are independent, continuous random variables, the Theil-Sen slope is a Fisher consistent slope estimator.
\end{theorem}

Before proving the validity of the Theil-Sen slope in the LiNGAM model we need the additional concept of correlation-faithfulness \citep{corrfaith}:

\begin{mydef}[Correlation-faithfulness]
    The distribution of $(X_1,\dots,X_p)$ is said to be correlation-faithful to the underlying graph if and only if the (conditional) correlations of the $X_i$'s are implicated by the graph structure.
\end{mydef}

Now suppose the data are realizations of a $p$-variate random vector $(X_1, \dots, X_p) \sim F_p$. Lemma \ref{lemma:generalizedlemma1} then states that the Theil-Sen slope can successfully identify exogenous variables and generalizes Lemma 1 of \cite{DirectLiNGAM}.

\begin{lemma}[Generalization of Lemma 1 of \cite{DirectLiNGAM}]\label{lemma:generalizedlemma1}
    Suppose that the random variables $X_1,\dots,X_p$ strictly follow the LiNGAM assumptions and that their distribution is correlation-faithful. We consider slope estimators as functionals $T$ acting on bivariate distributions $(X,Y) \sim F_2$. Assume that following properties hold for these slope functionals $T(F_2) = T(X,Y)$ when $Y$ is regressed on $X$:
    \begin{align}
    \label{eq:assumpslope}
        1. \quad  & T \text{ is regression equivariant: } \forall \gamma \in \mathbb{R}: T(X, Y + X \gamma) = T(X, Y) + \gamma \qquad \qquad \qquad \qquad \qquad \nonumber\\
        2. \quad &\text{If } X \text{and } Y \text{are independent:} \ T(F_2) = T(X, Y) = 0 \qquad \\
        3. \quad  & T(X,Y) \neq 0 \implies T(Y,X) \neq 0 \nonumber
    \end{align}
    \noindent
    Define the residual when $X_i$ is regressed on $X_j$ using slope functional $T$ as the following random variable: $r_i^{(j)} \coloneqq X_i - T(X_j,X_i) X_j \ (i \neq j)$. Then the variable $X_j$ is exogenous if and only if $X_j$ is independent of $r_i^{(j)}$ for all $i \neq j$. In particular, this holds for the Theil-Sen slope.
\end{lemma}

Next, Lemma \ref{lemma:generalizedlemma2} shows that the LiNGAM model holds on the residuals after an exogenous variable is regressed out using the Theil-Sen slope.

\begin{lemma}[Generalization of Lemma 2 of \cite{DirectLiNGAM}]\label{lemma:generalizedlemma2}
    Suppose that the random variables $X = (X_1,\dots,X_p)^T$ strictly follow the LiNGAM assumptions and that their distribution is correlation-faithful. Assume that the variable $X_j$ is exogenous and denote by $r^{(j)}$ the $(p-1)$-dimensional vector holding all the residuals when the $X_i, \ i \neq j,$ are regressed on $X_j$ using the a slope estimator satisfying the properties in \eqref{eq:assumpslope}. Then a LiNGAM holds for the residual random variables $r^{(j)}: \ r^{(j)} = B^{(j)} r^{(j)} + e^{(j)}$ or $r^{(j)} = A^{(j)} e^{(j)}$. Moreover, the causal order is preserved: $k_{r^{(j)}}(l) < k_{r^{(j)}}(m) \iff k(l) < k(m)$.
\end{lemma}

We conclude that using the Theil-Sen slope is effective at identifying exogenous variables and regressing out their effect, and hence our new method correctly identifies the underlying causal model under the LiNGAM assumptions. We will refer with TSLiNGAM (Theil-Sen LiNGAM) to the resulting algorithm which uses Theil-Sen regression for identifying exogenous variables.

\subsection{Robustness}

Theil-Sen regression is not only more efficient than OLS at heavy-tailed and skewed distributions, it is also more robust in the sense that it is more resilient against contamination in the data. As discussed, the Theil-Sen slope has a bounded influence function \citep{hampel1986robust}, implying that the effect that a single outlying observation can have on the measure is limited. Furthermore, the considered slope has a breakdown value of 0.293, meaning that the slope is robust up to 29.3\% contamination in the data \citep{rousseeuw2005robust}. In contrast, the OLS has an unbounded influence function and a breakdown value of 0\%. To further explore the effect of the robustness-efficiency trade-off, we additionally consider the repeated median, defined by \cite{Siegel}, for the identification of the exogenous variables. The repeated median is defined as: 

\begin{mydef}[Repeated median slope]
    For the linear regression of a random variable $Y$ on $X$, $Y = \beta X + e$, the repeated median slope estimator is defined as
\begin{equation}
    \label{eq:RepMed}
    \hat{\beta} = \text{med}_i \text{med}_{j \neq i} \frac{y_i-y_j}{x_i - x_j}, \qquad \text{for } x_j \neq x_i
\end{equation}
for data pairs $\{ (x_i,y_i) : i = 1, \dots,n \}$.
\end{mydef}
The functional form of the repeated median is given by:
\begin{align}
\label{eq:repmedfunctional}
    T(X,Y) &= \underset{X,Y}{\med} \left( \underset{X',Y'}{\med}\left( \frac{Y - Y'}{X - X'} \right) \right) \\
     \text{ with } Y &\mydistr Y', \ X \mydistr X' \nonumber
\end{align}

Just as the Theil-Sen slope, the repeated median is unbiased and regression equivariant. However, the repeated median is more robust, with a bounded influence function and a breakdown value of 0.5. In order to use the repeated median for the identification of exogenous variables, we need to verify its plug-in theoretically. As the Fisher consistency of the repeated median has only been proven for symmetric errors, see \cite{Siegel}, we proof this property for general error distributions.

\begin{theorem}[Fisher consistency of the repeated median]
\label{thm:RMfishcons}
    For a simple linear regression model $Y = \beta X + \varepsilon$ such that $X$ and $\varepsilon$ are independent, continuous random variables, the repeated median slope estimator is Fisher consistent.
\end{theorem}

As the repeated median is regression equivariant, Fisher consistent and makes use of medians, Lemma \ref{lemma:generalizedlemma1} and \ref{lemma:generalizedlemma2} also hold for this slope estimator. Hence it can thus also be used to identify exogenous variables in a LiNGAM structure.\\

\begin{figure}[h!]
\centering
\includegraphics[trim={0.3cm 0cm 2.5cm 0cm},clip,width =  0.23 \textwidth]{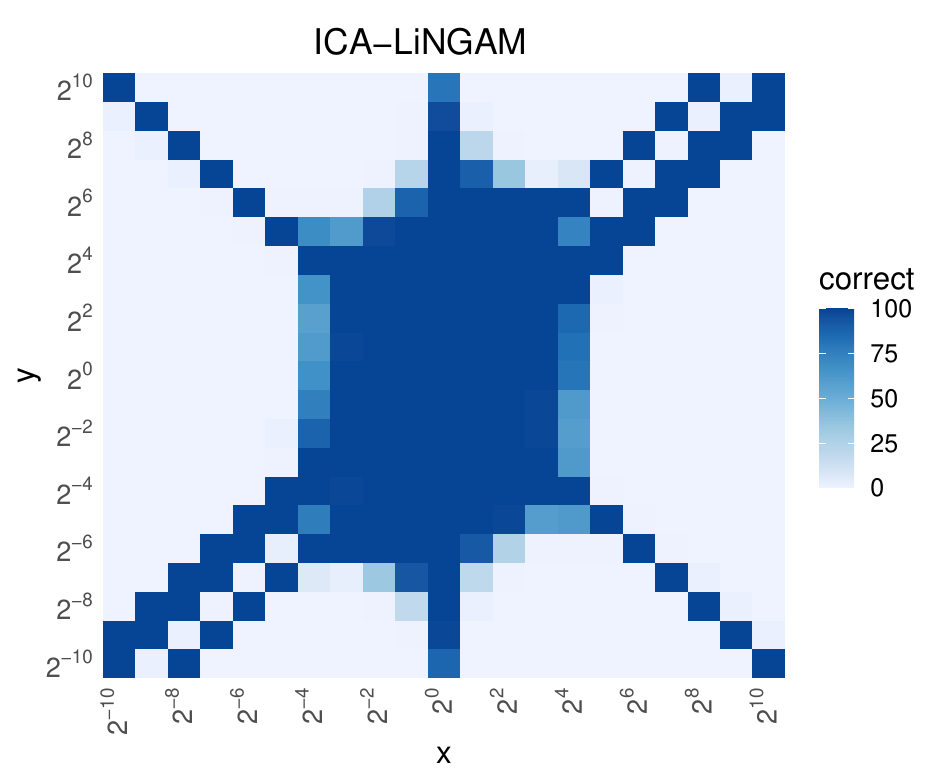}
\includegraphics[trim={0.3cm 0cm 2.5cm 0cm},clip,width =  0.23 \textwidth]{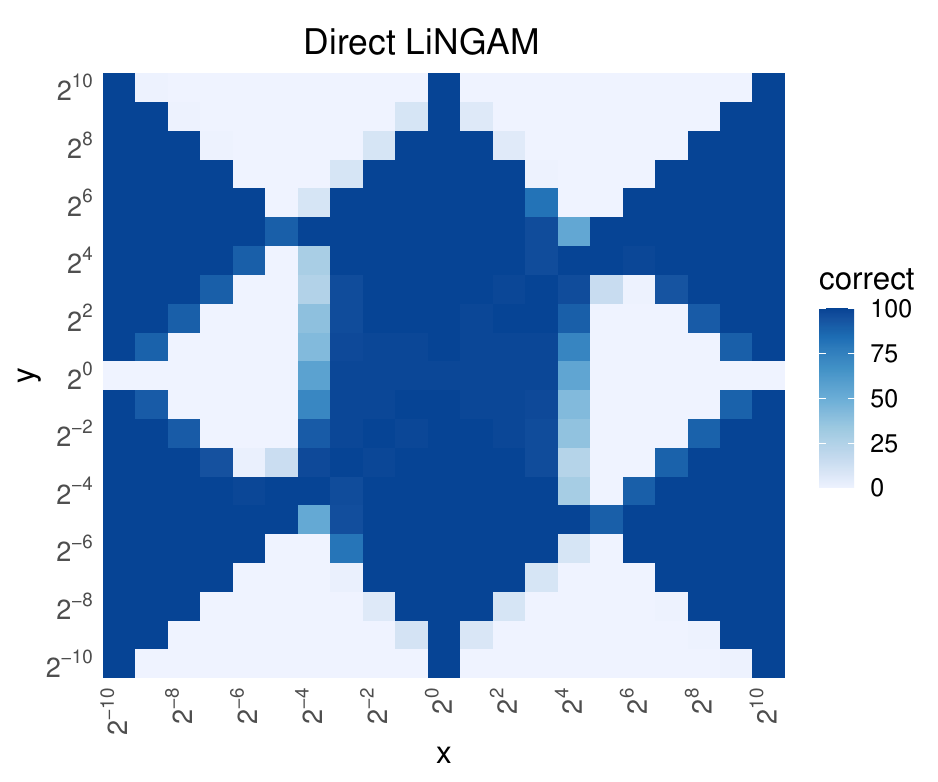} 
\includegraphics[trim={0.3cm 0cm 2.5cm 0cm},clip,width =  0.23 \textwidth]{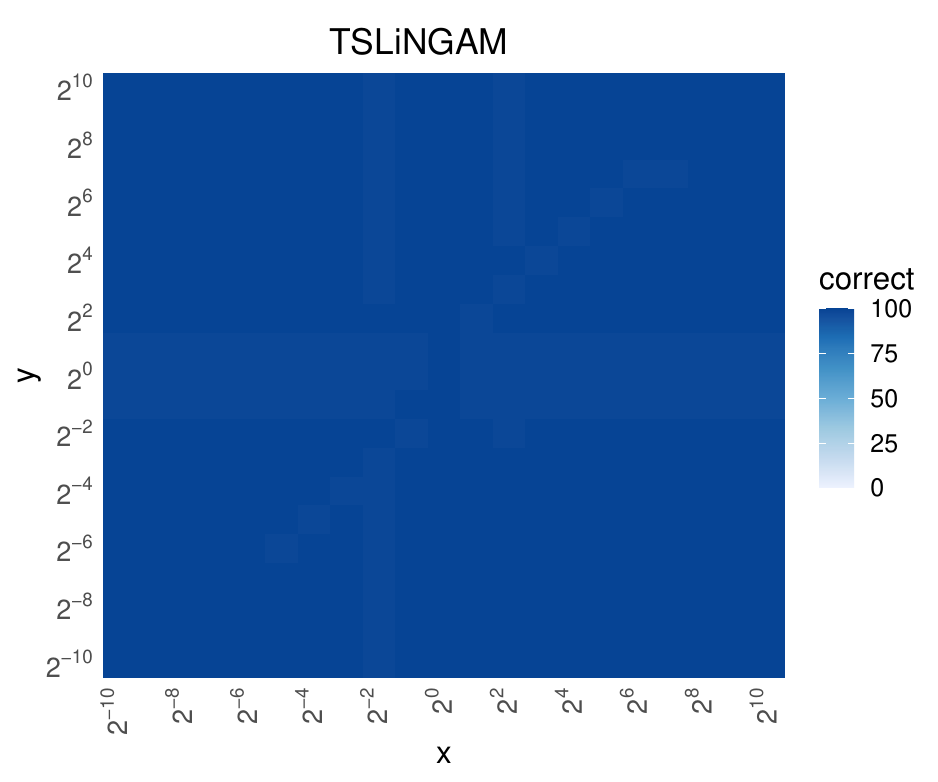}
\includegraphics[trim={0.3cm 0cm 0.2cm 0cm},clip,width =  0.27 \textwidth]{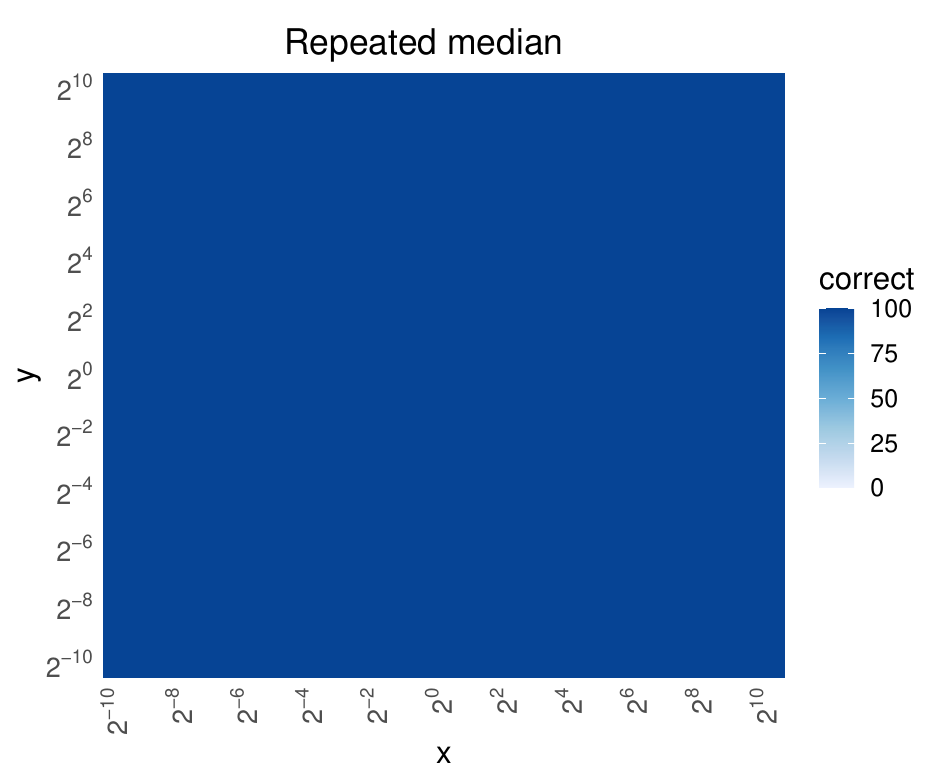}
\caption{\centering  Number of correctly estimated causal orders out of 100 runs per discovery method given the ($x,y$)-coordinate of added outlier.}
\label{fig:singleoutlier}
\end{figure}

We now illustrate the effect that a single outlier can have on LiNGAM methods in the following robustness experiment. The objective is to estimate the causal order of a simple two node DAG in the LiNGAM family:
\begin{equation*}
    \begin{cases}
        X_1 = e_1\\
        X_2 = X_1 + e_2
    \end{cases}
\end{equation*}
Here $e_1$ and $e_2$ are distributed according to a Student-$t$ distribution with 5 degrees of freedom. We generate $n=500$ observations of this bivariate causal model and replace one observation by an outlier of value $(\pm 2^i, \pm 2^j)$ for $i,j \in \{0,1,2,\dots,10\}$. The causal direction is then estimated from the contaminated data using the original ICA-LiNGAM algorithm from the \textit{pcalg} package \citep{pcalg}, the DirectLiNGAM algorithm and our adapted versions using Theil-Sen and the repeated median. We iterate this process 100 times to get representative results. The outcome is shown in Figure \ref{fig:singleoutlier}. For ICA-LiNGAM and DirectLiNGAM, we observe that a single outlying observation has the ability to distort the discovery of the causal order, even in the simplest of causal models. Both plots show different regions for the outlier such that the obtained causal order is the inverse of the true causal direction. In contrast, when Theil-Sen (TSLiNGAM) or the repeated median are used to identify the exogenous variables, we notice that the causal order is always correctly estimated, regardless of the values of the added outlier.

According to Figure \ref{fig:singleoutlier}, using the repeated median to identify the exogenous variables seems promising. As discussed, it is a robust slope estimator, but at a cost of lower efficiency. Notwithstanding, the repeated median is still more efficient than OLS for various error distributions, e.g. skewed, heavy-tailed or discrete distributions. Therefore from now on, we will also consider the repeated median as an alternative slope estimator and we will compare its use to TSLiNGAM in an extensive simulation study in Section \ref{sec:simulation}.

\subsection{Computational considerations}
\label{sec:compcons}

Finally, we discuss the computational cost of DirectLiNGAM and the proposed TSLiNGAM. As Theil-Sen regression and the repeated median can be computed in $\mathcal{O}(n\log(n))$ time \citep{cole1989optimal, katz1993optimal, stein1992finding, matouvsek1993efficient}, they do not add much to the computational cost when used instead of simple OLS (which requires $\mathcal{O}(n)$ time). Therefore, TSLiNGAM has a computational cost that is similar to that of DirectLiNGAM.

The computational costs of both algorithms is in fact dominated by the independence measure for each remaining variable per iteration in the algorithm. In the original paper \citep{DirectLiNGAM}, the independence measure used in Equation \eqref{eq:KBIM} is the kernel-based estimator of mutual information defined by \cite{BachJordan}. Although this measure performs well, its accumulated computational cost can become rather large, as lots of Gram matrices with Gaussian kernels have to be computed.

Therefore, as an alternative independence measure, we consider the use of the distance correlation (dcorr) between random variables \citep{dcor}. This measure, unlike Pearson's correlation, is zero if and only if the variables are independent. One of the advantages is that it can be computed in $O(n \text{log} (n) )$ time \citep{fastdcor}. In contrast, the kernel-based independence measure computed on two variables has a computational complexity of $O(nM^2+M^3)$, where $M$ ($\ll n$) is the maximal rank found by the low-rank approximations of the Gram matrices used in the algorithm for the independence measure. Therefore, only if $M = \mathcal{O}(\sqrt{\log(n)})$, we obtain the same computational complexity as dcorr. Empirically, $M$ often seems to grow substantially faster than this rate. Hence, when the data
sets are too large to use the kernel-based independence measure, it can be beneficial to use the distance correlation instead to speed up DirectLiNGAM and TSLiNGAM.

\section{Simulation}
\label{sec:simulation}

In this section we compare the proposed method with direct competitors. 

\subsection{Setup and methods}

We compare TSLiNGAM with the original DirectLiNGAM algorithm and with extremal ancestral search (EASE) \citep{EASE}. The latter method is designed for heavy-tailed data, and is thus a highly relevant competitor to TSLiNGAM. In addition, we also compare with three variations of the proposed algorithm. The first uses TS regression with dcorr as independence measure. The other two use RM regression paired with KBI and dcorr respectively.

The data is generated by the following procedure, inspired by the simulation setup of \cite{EASE}:
\begin{enumerate}
    \item We simulate data of dimension $p = 2$ with sample sizes $n = \{ 5,10,25,50,100 \}$, of dimension $p = 5$ with sample sizes $\{ n = 30,50,100,200 \}$ and of dimension $p = 10$ with sample sizes $\{ n = 50,100,200,300 \}$.
    \item We generate a linear structural causal model $X = BX + e$ with $X = (X_1,\dots,X_p)^T$, $e = (e_1,\dots,e_p)^T$ and $B \in \mathbb{R}^{p \times p}$ as follows:
    \begin{enumerate}
        \item First, we generate a random causal order between the variables $X_1,\dots,X_p$ as a permutation $\pi$ of $\{1,\dots,p\}$.
        \item Per variable $X_i$ with $\pi (i) > 1$, the number of parents is distributed as $Bin(\pi (i) - 1, q)$ with $q=\{ 1, 0.6, 0.5 \}$ respectively for dimension $p=\{ 2,5,10 \}$.
        \item Next, we select those parents randomly from the variables with a lower causal order, such that cycles are ruled out.
        \item Then, we sample the connection strengths $B_{ij}$ per variable per parent uniformly from $[-0.9,-0.1] \cup [0.1,0.9]$. This yields the matrix $B$.
        \item Finally, we sample the noise variables randomly from following distributions: Student-$t$ with 1, 2 or 5 degrees of freedom, a centered lognormal distribution, a centered Pareto distribution and a centered exponential distribution. Combining $B$ and $e$, we obtain $X$.
    \end{enumerate}
\end{enumerate}
For each setting, we generate 1000 data sets. To compare performance among the different methods, we count the number of times the algorithm returns the correct causal order. All implementations are done in \texttt{R}. For the Theil-Sen slope and the repeated median we use the corresponding functions from the \textit{robslopes} package \citep{robslopes, RJ-2023-012}. For the distance correlation we use the \textit{dccpp} package \cite{dccpp} and for EASE we use the implementation in the \textit{causalXtreme} package \citep{causalXtreme}.

\subsection{Results}

We discuss the results for $p=10$ variables here. The results for $p=2$ and $p=5$ are qualitatively similar and can be found in Section \ref{app:simresults} of the Appendix. The results for $p=10$ are presented in Table \ref{tab:simulationp10}.

We observe that in almost every scenario the proposed TSLiNGAM achieves the best result. TSLiNGAM strongly outperforms DirectLiNGAM when the data is heavy-tailed. For skewed distributions, TSLiNGAM also performs better than DirectLiNGAM, although the difference is somewhat smaller. As the distribution moves closer to normality, such as for the $t_5$ distribution, DirectLiNGAM becomes the preferred method. However, note that the difference in performance is almost negligible and perhaps more importantly, the absolute performance is very poor. This is explained by the fact that the identifiability of the LiNGAM structure is lost when there are Gaussian errors, and as we move closer to that scenario, it becomes increasingly difficult to identify the underlying structure. EASE does not perform well here. This is probably explained by the fact that EASE only looks at the tails and therefore needs bigger sample sizes in order to perform well. Finally, we consider the variations of the TSLiNGAM algorithm. The repeated median performs well on heavy-tailed distributions, but does not offer an improvement over TSLiNGAM. Using dcorr as independence measure becomes a viable strategy when the sample size is reasonably large.

\begin{table}[h!]
\centering
\begin{tabular}{|r|rrrr|rrrr|}
\hline
 sample size & 50 & 100 & 200 & 300 & 50 & 100 & 200 & 300 \\ 
    \hline
  & \multicolumn{4}{c|}{$t_1$} &   \multicolumn{4}{c|}{Pareto } \\
  \hline
TSLiNGAM & \textbf{477} & \textbf{806} & \textbf{942} & \textbf{984} & \textbf{539} & \textbf{892} & \textbf{983} & \textbf{1000} \\ 
  DirectLiNGAM & 286 & 432 & 555 & 640 & 368 & 552 & 733 & 840 \\ 
  EASE & 21 & 124 & 307 & 450 & 3 & 10 & 17 & 27 \\ 
  Repeated Median \&  KBI & 379 & 733 & 915 & 969 & 365 & 759 & 961 & 981 \\ 
  Theil-Sen \&  dcorr & 232 & 530 & 747 & 841 & 361 & 740 & 942 & 982 \\ 
  Repeated Median \&  dcorr & 152 & 425 & 657 & 791 & 178 & 548 & 877 & 947 \\ 
    \hline
  & \multicolumn{4}{c|}{$t_2$} &   \multicolumn{4}{c|}{lognormal } \\
  \hline
  TSLiNGAM & \textbf{167} & \textbf{387} & \textbf{601} & 711 & \textbf{457} & \textbf{798} & \textbf{959} & \textbf{992} \\ 
  DirectLiNGAM & 149 & 333 & 539 & 642 & 351 & 649 & 859 & 924 \\ 
  EASE & 3 & 13 & 52 & 81 & 0 & 3 & 8 & 12 \\ 
  Repeated Median  \& KBI & 128 & 348 & 592 & 663 & 309 & 675 & 919 & 981 \\ 
  Theil-Sen  \& dcorr & 38 & 189 & 519 & \textbf{733} & 273 & 703 & 932 & 980 \\ 
  Repeated Median  \& dcorr & 21 & 133 & 427 & 651 & 156 & 514 & 848 & 937 \\ 
    \hline
  & \multicolumn{4}{c|}{$t_5$} &   \multicolumn{4}{c|}{exponential } \\
  \hline
  TSLiNGAM & \textbf{17} & 35 & 88 & 151 & 236 & \textbf{601} & 858 & \textbf{942} \\ 
  DirectLiNGAM & 16 & \textbf{37} & \textbf{90} & \textbf{158} & \textbf{245} & 575 & 812 & 903 \\ 
  EASE & 0 & 1 & 5 & 1 & 0 & 0 & 1 & 1 \\ 
  Repeated Median \&  KBI & 11 & 34 & 72 & 132 & 155 & 482 & 818 & 896 \\ 
  Theil-Sen \&  dcorr & 2 & 1 & 7 & 11 & 140 & 542 & \textbf{867} & 941 \\ 
  Repeated Median \&  dcorr & 0 & 2 & 4 & 16 & 66 & 360 & 746 & 867 \\ 
   \hline  
\end{tabular}
\caption{Number of correct causal orders out of 1000 trials for p = 10.}
\label{tab:simulationp10}
\end{table}

\subsubsection{Computation time}

In addition to the results on the recovery of the underlying LiNGAM structure, we study the computation times of the methods. We discuss the computation times for the simulation study with $p=10$ and for the $t_5$ distribution here. The other distributions had similar computational costs. The computation times for $p=2$ and $p=5$ are qualitatively similar and can be found in Section \ref{app:simresults} of the Appendix.

 Table \ref{tab:comptimep10} presents the computation times for $p=10$ and the $t_5$ distribution. The first thing to note is that TSLiNGAM and its variants have essentially the same computational cost as DirectLiNGAM. This is explained by the fact that the computation time of both algorithms is dominated by the kernel-based independence measure. As a result, when using dcorr as a measure of independence, we see a substantial speedup of about one order of magnitude. This suggests that dcorr is useful when the sample size gets larger, which is precisely the scenario in which its performance is also competitive with TSLiNGAM. Finally, we note that EASE is by far the fastest method here. However, as pointed out before, it is not competitive in these relatively small-sample scenarios.

\begin{table}[h!]
\centering
\begin{tabular}{|r|rrrr|}
 \hline
 sample size & 50 & 100 & 200 & 300 \\ 
  \hline
TSLiNGAM & 15.68 & 19.32 & 24.68 & 30.19 \\ 
  DirectLiNGAM & 14.64 & 17.98 & 22.90 & 27.28 \\ 
  EASE & 0.03 & 0.03 & 0.04 & 0.04 \\ 
  Repeated Median \& KBI & 15.98 & 19.83 & 26.05 & 32.83 \\ 
  Theil-Sen \& dcorr & 0.96 & 1.54 & 2.62 & 3.62 \\ 
  Repeated Median \& dcorr & 1.45 & 2.64 & 4.43 & 6.61 \\ 
   \hline
\end{tabular}
\caption{Computational time in minutes for 1000 runs for $p=10$ for a Student-$t$ distribution with 5 degrees of freedom.}

\label{tab:comptimep10}
\end{table}

\section{Real data applications}
\label{sec:realdata}
In this section we illustrate the application of TSLiNGAM on four data sets from medical and social sciences.

\subsection{Physician data}

As a first real data example, we consider data originating from the US National Medical Expenditure Survey conducted in 1987 and 1988. This data contains health-related information on 4406 individuals and can be found at \href{http://qed.econ.queensu.ca/jae/1997-v12.3/deb-trivedi/}{http://qed.econ.queensu.ca/jae/1997-v12.3/deb-trivedi/} or in the R-package \textit{AER} \citep{AER} as the data set NMES1988.  We work with the following variables: age, school (years of education), income (family income), chronic (number of chronic conditions), visits (number of physician office visits) and hospital (number of hospital stays). 

\begin{figure}[h!]
    \centering
        \resizebox{\columnwidth}{9cm}{
    \begin{tikzpicture}
        \begin{scope}[every node/.style={circle,thick,draw}]
            \node[shape=rectangle,draw=black,align=center] (1) at (0,0) {Hospital};
            \node[shape=rectangle,draw=black,align=center] (2) at (-1.5,-1.5) {Chronic};
            \node[shape=rectangle,draw=black,align=center] (3) at (1.5,-3) {Visits};
            \node[shape=rectangle,draw=black,align=center] (4) at (0,-4.5) {Age};
            \node[shape=rectangle,draw=black,align=center] (5) at (-1.5,-6) {Income};
            \node[shape=rectangle,draw=black,align=center] (6) at (1.5,-7.5) {School};

            \node[shape=rectangle,draw=black,align=center] (7) at (8,0) {Age};
            \node[shape=rectangle,draw=black,align=center] (8) at (6,-1.5) {School};
            \node[shape=rectangle,draw=black,align=center] (9) at (9.5,-3) {Income};
            \node[shape=rectangle,draw=black,align=center] (10) at (6.7,-4.5) {Chronic};
            \node[shape=rectangle,draw=black,align=center] (11) at (10,-6) {Visits};
            \node[shape=rectangle,draw=black,align=center] (12) at (8,-7.5) {Hospital};
        \end{scope}

        \begin{scope}[>={Stealth[black]},
              every edge/.style={draw,very thick,}]
            \path [->] (1) edge node {} (2);
            \path [->] (1) edge node {} (3);
            \path [->] (1) edge [bend right=10] node {} (4);
            \path [->] (1) edge [bend left=50] node {} (6);
            \path [->] (2) edge node {} (3);
            \path [->] (2) edge node {} (4);
            \path [->] (2) edge [bend right=30] node {} (6);
            \path [->] (3) edge node {} (4);
            \path [->] (3) edge node {} (6);
            \path [->] (4) edge node {} (5);
            \path [->] (4) edge node {} (6);
            \path [->] (5) edge node {} (6);

            \path [->] (7) edge node {} (8);
            \path [->] (7) edge node {} (9);
            \path [->] (7) edge node {} (10);
            \path [->] (7) edge [bend left=50] node {} (11);
            \path [->] (7) edge node {} (12);
            \path [->] (8) edge node {} (9);
            \path [->] (8) edge node {} (11);
            \path [->] (8) edge [bend right=40] node {} (12);
            \path [->] (10) edge node {} (11);
            \path [->] (10) edge node {} (12);
            \path [->] (11) edge node {} (12);
        \end{scope}
    \end{tikzpicture}
    }
    \caption{Directed acyclic graphs found by DirectLiNGAM (left) and TSLiNGAM (right).}
    \label{fig:physicianvisits}
\end{figure}
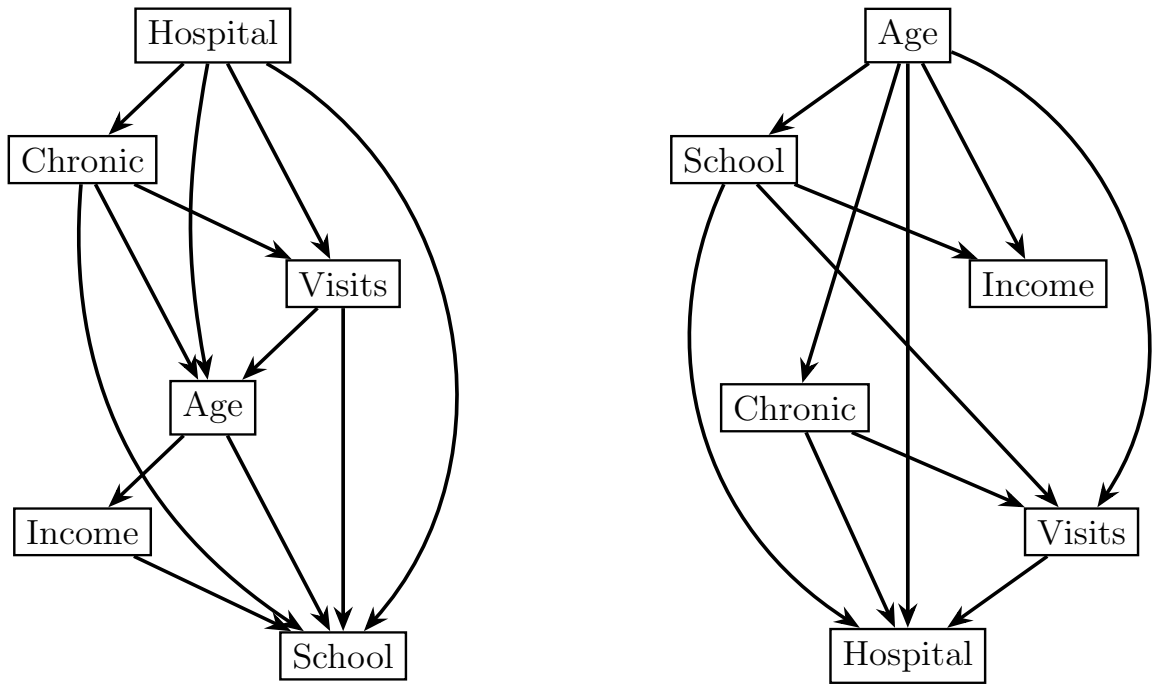

We compare TSLiNGAM with the standard DirectLingam to find the causal structure. To prune redundant edges in the resulting adjacency matrices $B$, we perform Adaptive Lasso, as is done in \cite{DirectLiNGAM}. This results in the directed acyclic graphs shown in Figure \ref{fig:physicianvisits}.

The causal order found by DirectLiNGAM is (hospital, chronic, visits, age, income, school). This order is not very logical. We would for example expect that the number of chronic conditions and a person's age are causes of the number of hospital stays. Additionally, years of schooling should have an impact on a persons income. In contrast, the causal order found by TSLiNGAM is (age, school, income, chronic, visits, hospital). This order is very logical and corresponds with our intuition. Furthermore, the causal graph found by TSLiNGAM consists of edges that match our understanding of the variables.

\begin{figure}[h!]
\centering
\includegraphics[trim={.5cm .5cm .5cm .5cm},clip,height=7cm]{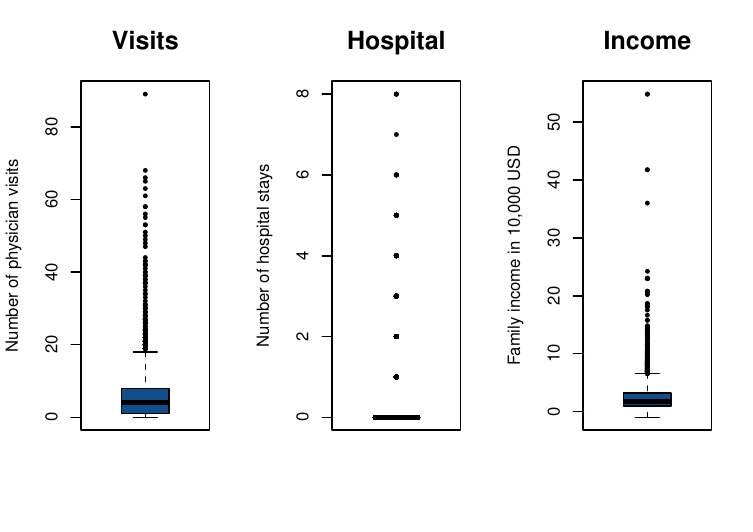}
\caption{Boxplots of the heavy-tailed variables in the physician data set.}
\label{fig:boxplots_physiciandata}
\end{figure}

The better result obtained by TSLiNGAM can be explained by studying the underlying variables of the data set. We know that TSLiNGAM tends to outperform DirectLiNGAM on heavy-tailed and skewed data, and it turns out that the variables visits, hospital and income are leptokurtic and have very fat tails, see for instance the boxplots in Figure \ref{fig:boxplots_physiciandata}. 

\subsection{GAGurine data}
The second data set considered in this work is the GAGurine data from the package \textit{MASS} in R \citep{MASS}. This data contains the concentration of the chemical GAG in the urine of 314 children between 0 and 17 years old, where it is known that age is a dominant cause of the concentration of GAG. 

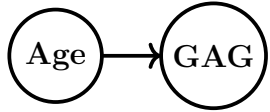
\begin{figure}[h!]
    \centering
    \begin{tikzpicture}
      \SetGraphUnit{2.1}
      \Vertex{GAG}
      \WE(GAG){Age}
      \Edge[](Age)(GAG)
    \end{tikzpicture}
    \caption{Ground truth: age is a cause of GAG concentration in urine.}
\end{figure}

\begin{figure}[h!]
    \centering
    \includegraphics[trim={0cm 0cm 0cm 0cm},clip,height=8.5cm]{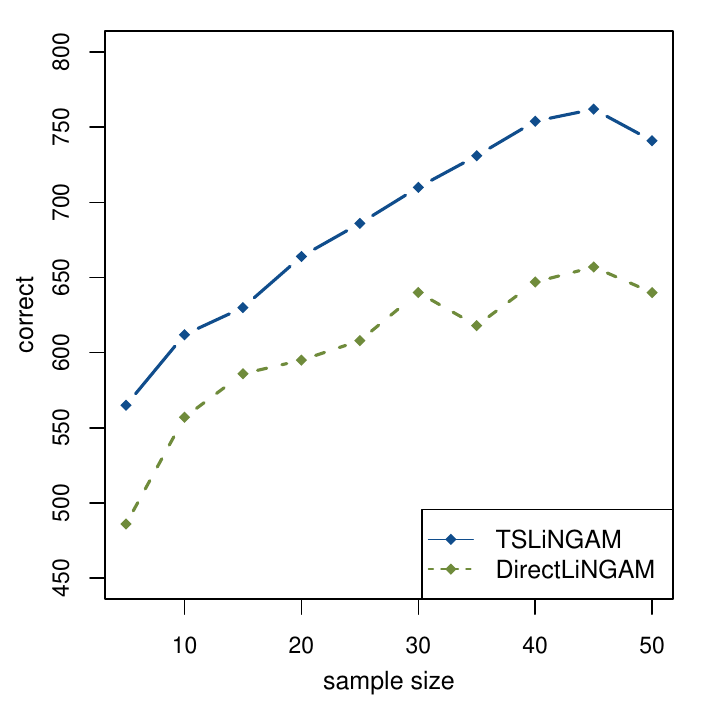}
    \captionsetup{width=0.9\textwidth}
    \caption{Number of correctly found causal orders by TSLiNGAM and DirectLiNGAM for 1000 runs on subsamples of the GAGurine data with specified sample sizes.}
    \label{fig:GAGurine}
\end{figure}

\noindent
On this data set, both DirectLiNGAM and TSLiNGAM succeed in discovering the right causal order. However, as we would like to demonstrate the small-sample efficiency of TSLiNGAM, we sample 1000 data subsets of sizes $\{5,10,15,\dots,50\}$ from the original data. The number of times TSLiNGAM and DirectLiNGAM find the right causal order on these subsamples are shown in Figure \ref{fig:GAGurine}. Overall we observe that TSLiNGAM has a 10\% higher small-sample efficiency compared to DirectLiNGAM. For example, for sample size $n=45$, TSLiNGAM finds the right causal order 762 times, while DirectLiNGAM only succeeds 657 times. This increase in efficiency can be explained by observing that the distribution of the variable GAG is right-skewed and tailed, a scenario where the Theil-Sen slope is better suited than OLS.

\subsection{FMRI data}
\begin{figure}[h!]
        \centering
        \begin{tikzpicture}[rotate=90]
            \SetGraphUnit{1.8}
            \Vertices{circle}{1,2,3,4,5}
            \Edge[](1)(2)
            \Edge[](2)(3)
            \Edge[](3)(4)
            \Edge[](4)(5)
            \Edge[](1)(5)
        \end{tikzpicture}
        \caption{The causal order of the FMRI data is (1,2,3,4,5).}
        \label{fig:FMRI}
    \end{figure}
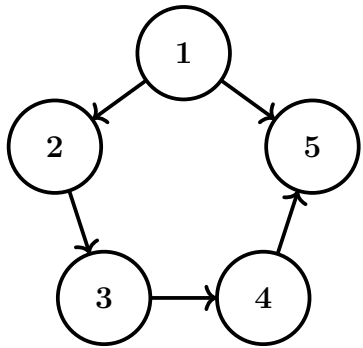

As a third data set, we study the functional magnetic resonance imaging (FMRI) data simulated in \cite{SMITH}. This data was previously studied within the LiNGAM framework \citep{SMITH,PWDL,parceLiNGAM}, however we now use the FMRI data in a robustness context. We take the first simulation data set from the paper which contains 10.000 continuous observations of 5 variables and has a causal structure as demonstrated in Figure \ref{fig:FMRI}.

\noindent
If we run DirectLiNGAM or TSLiNGAM on the original data set, both methods succeed in discovering the correct causal order. However, to demonstrate the robustness of TSLiNGAM, we artificially contaminate 20 observations by changing the value of the first variable to a number generated from a Student-$t$ distribution with 3 degrees of freedom, centered around 25. We then run both algorithms 100 times again and observe that DirectLiNGAM never finds the right causal order. TSLiNGAM, in contrast, recovers the ground truth 72 times and is therefore much less influenced by the 20 outlying observations. This shows that TSLiNGAM using the Theil-Sen slope instead of OLS is more resilient towards contamination.

\subsection{Sociological data}

As a final real data example, we try to replicate the example performed in the original DirectLiNGAM paper on sociology data. 
The data are publicly available at the General Social Survey (GSS: \href{https://gssdataexplorer.norc.org/gss_data}{https://gssdataexplorer.norc.org/gss\_data}). We study the variables father's occupation level, son's income, father's education, son's occupation level, son's education and number of siblings. We take the same subset of the data as studied by \cite{DirectLiNGAM}: non-farm background, ages 35 to 44, white, male, in the labor force at the time of the survey and years 1972 to 2006, see Table \ref{tab:GSSvariables} for details. After omitting observations which contain missing values, this results in a data set with 2117 observations. 

\begin{table}[h!]
\small
\begin{center}
\begin{tabular}{| c | c | c || c | c |}  
\hline 
& studied variables & GSS codebook name & selection variables & GSS codebook name\\
\hline
$X_1$ &  Father's occupation level & PAPRES10 & non-farm background & RES16\\
$X_2$ & Son's income & REALRINC & age & AGE\\
$X_3$ & Father's education & PAEDUC & white & RACE \\
$X_4$ & Son's occupation level & PRESTG10 & sex & SEX \\
$X_5$ & Son's education & EDUC & in the labor force & WRKSTAT\\
$X_6$ & Number of siblings & SIBS & year & YEAR\\
 \hline 
\end{tabular}
\end{center}
\captionsetup{width=0.85\textwidth}
\caption{Studied variables taken from the GSS repository and which variables we selected our sample on.}
\label{tab:GSSvariables}
\end{table}

\begin{figure}[h!]
    \centering
    \begin{tikzpicture}
        \begin{scope}[every node/.style={circle,thick,draw}]
            \node[shape=rectangle,draw=black,align=center] (A) at (0,0) {Father's education\\
            $(X_3)$};
            \node[shape=rectangle,draw=black,align=center] (B) at (0,-2.25) {Father's occupation\\
            $(X_1)$};
            \node[shape=rectangle,draw=black,align=center] (C) at (0,-4.5) {Number of siblings\\
            $(X_6)$};
            \node[shape=rectangle,draw=black,align=center] (D) at (5,-1) {Son's education\\
            $(X_5)$};
            \node[shape=rectangle,draw=black,align=center] (E) at (5,-3) {Son's occupation\\
            $(X_4)$};
            \node[shape=rectangle,draw=black,align=center] (F) at (10,-2) {Son's income\\
            $(X_2)$};
        \end{scope}
        \begin{scope}[>={Stealth[black]},
              every edge/.style={draw,very thick}]
            \path [{Stealth[color=violet]}-{Stealth[color=violet]},draw=violet] (A) edge node {} (B);
            \path [{Stealth[color=violet]}-{Stealth[color=violet]},draw=violet] (B) edge node {} (C);
            \path [{Stealth[color=violet]}-{Stealth[color=violet]},draw=violet] (A) edge[bend right=65] node {} (C);
            \path [->] (A) edge node {} (D);
            \path [->] (B) edge node {} (D);
            \path [->] (B) edge node {} (E);
            \path [->] (C) edge node {} (D);
            \path [->] (C) edge node {} (E);
            \path [->] (D) edge node {} (E); 
            \path [->] (D) edge node {} (F);
            \path [->] (E) edge node {} (F);
        \end{scope}
    \end{tikzpicture}
    \captionsetup{width=0.85\textwidth}
    \caption{Ground truth based on domain knowledge: a directed arrow indicates a possible causal relation, a bidirected purple arrow signifies an unknown causal relation (there might be a relation, a latent confounder or nothing).}
    \label{fig:socdomknow}
\end{figure}
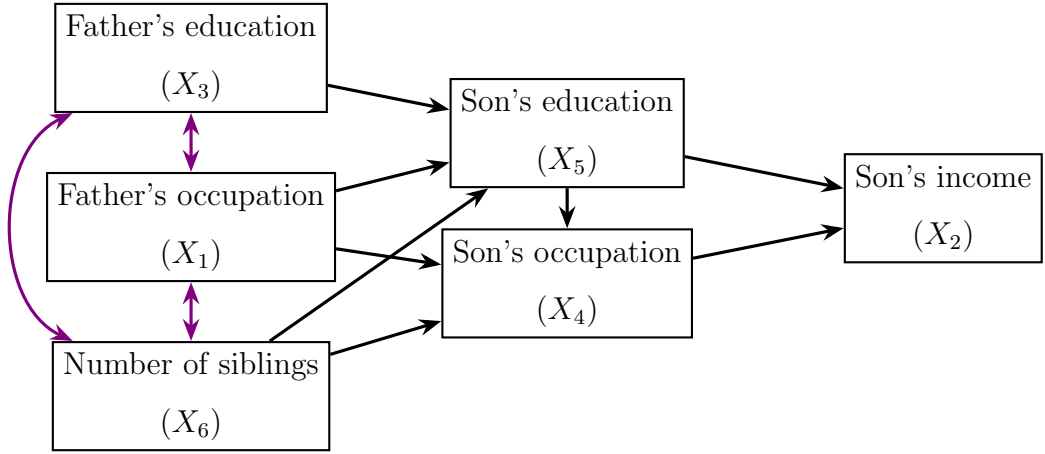

Domain knowledge on the causal relations between these variables suggests the causal structure shown in Figure \ref{fig:socdomknow} \citep{DirectLiNGAM}.

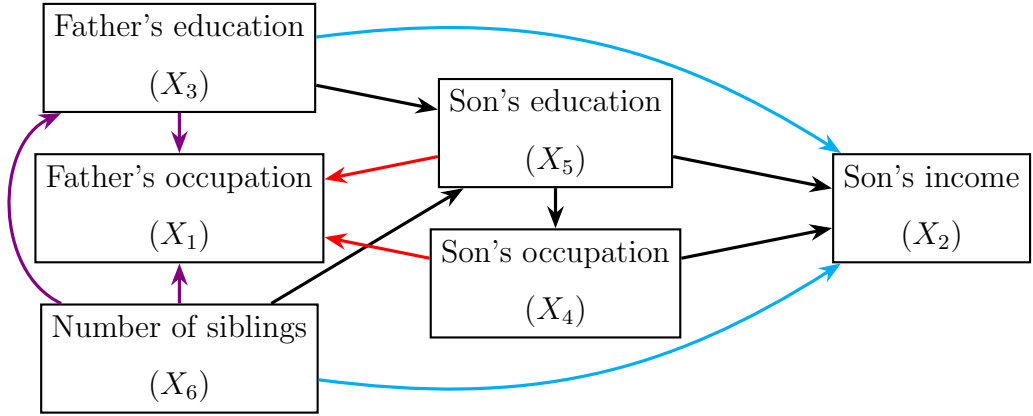
\begin{figure}[h!]
    \centering
    \begin{tikzpicture}
        \begin{scope}[every node/.style={circle,thick,draw}]
            \node[shape=rectangle,draw=black,align=center] (3) at (0,0) {Father's education\\
            $(X_3)$};
            \node[shape=rectangle,draw=black,align=center] (1) at (0,-2) {Father's occupation\\
            $(X_1)$};
            \node[shape=rectangle,draw=black,align=center] (6) at (0,-4) {Number of siblings\\
            $(X_6)$};
            \node[shape=rectangle,draw=black,align=center] (5) at (5,-1) {Son's education\\
            $(X_5)$};
            \node[shape=rectangle,draw=black,align=center] (4) at (5,-3) {Son's occupation\\
            $(X_4)$};
            \node[shape=rectangle,draw=black,align=center] (2) at (10,-2) {Son's income\\
            $(X_2)$};
        \end{scope}

        \begin{scope}[>={Stealth[black]},
              every edge/.style={draw,very thick,}]
              \path [-{Stealth[color=violet]},draw=violet] (3) edge node {} (1);
              \path [->] (3) edge node {} (5);
              \path [->] (4) edge node {} (2);
              \path [->] (5) edge node {} (2);
              \path [->] (5) edge node {} (4);
              \path [-{Stealth[color=violet]},draw=violet] (6) edge node {} (1);
              \path [->] (6) edge node {} (5);
              \path [-{Stealth[color=violet]},draw=violet] (6) edge[bend left=65] node {} (3);
                            
              \path [-{Stealth[color=cyan]},draw=cyan] (6) edge[bend right=20] node {} (2);
              \path [-{Stealth[color=cyan]},draw=cyan] (3) edge[bend left=20] node {} (2);

              \path [-{Stealth[color=red]},draw=red] (5) edge node {} (1);
              \path [-{Stealth[color=red]},draw=red] (4) edge node {} (1);
              

        \end{scope}
    \end{tikzpicture}
    \captionsetup{width=0.85\textwidth}
    \caption{Causal graph found by DirectLiNGAM with correct arrows in black, wrongly directed arrows in red, unverifiable arrows in purple and redundant arrows in blue.}
    \label{fig:DAGDLsoc}
\end{figure}

\begin{figure}[h!]
    \centering
    \begin{tikzpicture}
        \begin{scope}[every node/.style={circle,thick,draw}]
            \node[shape=rectangle,draw=black,align=center] (3) at (0,0) {Father's education\\
            $(X_3)$};
            \node[shape=rectangle,draw=black,align=center] (1) at (0,-2) {Father's occupation\\
            $(X_1)$};
            \node[shape=rectangle,draw=black,align=center] (6) at (0,-4) {Number of siblings\\
            $(X_6)$};
            \node[shape=rectangle,draw=black,align=center] (5) at (5,-1) {Son's education\\
            $(X_5)$};
            \node[shape=rectangle,draw=black,align=center] (4) at (5,-3) {Son's occupation\\
            $(X_4)$};
            \node[shape=rectangle,draw=black,align=center] (2) at (10,-2) {Son's income\\
            $(X_2)$};
        \end{scope}

        \begin{scope}[>={Stealth[black]},
              every edge/.style={draw,very thick,}]
              \path [-{Stealth[color=violet]},draw=violet] (3) edge node {} (1);
              \path [->] (3) edge node {} (5);
              \path [->] (5) edge node {} (2);
              \path [->] (5) edge node {} (4);
              \path [->] (6) edge node {} (5);
              \path [-{Stealth[color=violet]},draw=violet] (6) edge node {} (1);
              \path [-{Stealth[color=violet]},draw=violet] (6) edge[bend left=65] node {} (3);
              
              \path [-{Stealth[color=red]},draw=red] (5) edge node {} (1);
              

        \end{scope}
    \end{tikzpicture}
    \captionsetup{width=0.85\textwidth}
    \caption{Causal graph found by TSLiNGAM with correct arrows in black, wrongly directed arrows in red and unverifiable arrows in purple (there are no redundant arrows).}
    \label{fig:DAGTSsoc}
\end{figure}
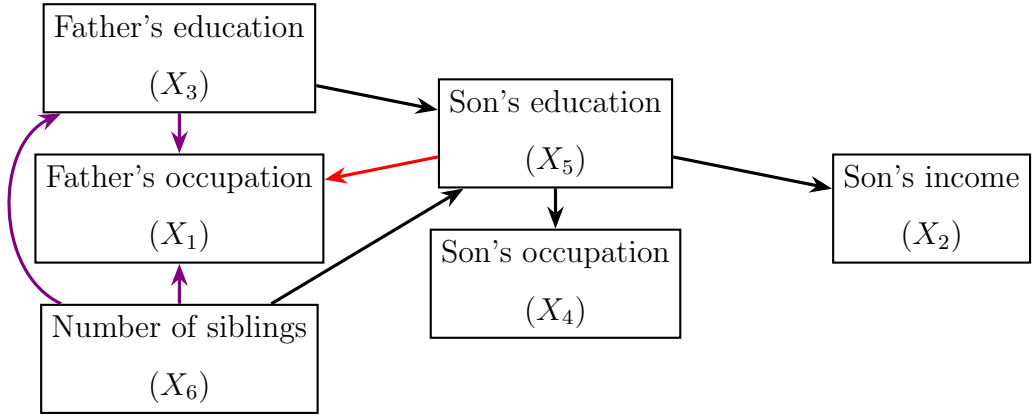

On this data we then run the DirectLiNGAM algorithm and our TSLiNGAM. To prune redundant edges in the resulting adjacency matrices $B$, we again perform Adaptive Lasso. This yields the directed acyclic graphs shown in Figures \ref{fig:DAGDLsoc} and \ref{fig:DAGTSsoc}. The two discovered DAGs are fairly good. DirectLiNGAM finds 5 correct edges, 2 wrongly directed edges and 2 redundant edges. TSLiNGAM finds 4 correct edges, 1 wrongly directed edge and no redundant edges. Overall, both methods perform equally well on the sociological data.

Additionally we remark that the Theil-Sen estimator combined with the distance correlation gave a better outcome, see the resulting DAG in Figure \ref{fig:DAGTSdcorsoc} of the Appendix. This combination of slope estimator and independence measure yields 6 correct edges, 1 wrongly directed edge and 2 redundant edges, the best result yet.

\section{Conclusion}\label{sec:conclusion}
In this work, we proposed TSLiNGAM, which builds on the popular DirectLiNGAM algorithm for causal discovery in LiNGAM structures. We proved that TSLiNGAM recovers the underlying causal LiNGAM structure by building on a new Fisher consistency result of the Theil-Sen slope.
By leveraging the attractive properties of the Theil-Sen slope estimator, we obtain improved recovery under heavy tailed and skewed data models, without sacrificing performance in models with symmetric distributions which are close to normal. This improved performance was illustrated in an extensive simulation study. 
Furthermore, we suggested considering a different independence measure in the algorithm. More precisely, using the distance correlation instead of the original kernel-based independence measure reduces the overall computational cost of the method without sacrificing performance, provided the data set is large enough.

We additionally illustrated the TSLiNGAM on four real data sets. These applications confirm better performance under skewness and heavy tails, an improved small-sample efficiency and increased robustness to outliers compared to the original DirectLiNGAM algorithm.

In summary, we conclude that the newly developed method can be considered a premium alternative to DirectLiNGAM. TSLiNGAM performs significantly better on heavy-tailed data and discovers the right causal order on smaller sample sizes, without increasing the computational cost. In addition, TSLiNGAM also showed better robustness properties as it is more resilient to contamination.

\section*{Disclosure statement}
The authors report there are no competing interests to declare.

\bigskip
\begin{center}
{\large\bf SUPPLEMENTARY MATERIAL}
\end{center}

\begin{description}

\item[appendix:] Document containing proofs, additional simulation results and figures. (.pdf file)

\end{description}

\bibliographystyle{abbrvnat}
\bibliography{TSLiNGAM_arxiv}

\clearpage
\appendix

\section{Proofs}\label{app:proofs}
\subsection{Proof of Lemma \ref{lemma:generalizedlemma1}}

\begin{proof}
$ $\newline
\noindent
\underline{1) Assume $X_j$ is exogenous:}\\
If $X_j$ is exogenous we have $X_j = e_j$. From $X = Ae$ we have $X_i = a_{ij} X_j + \sum_{h \neq j} a_{ih}e_h \ (i \neq j)$. Here $X_j$ is independent of $\sum_{h \neq j} a_{ih}e_h$ since $X_j = e_j$ and all the $e_i$ are mutually independent. Then:
\begin{align*}
    r_i^{(j)} &= X_i - T(X_j,X_i) X_j\\
    &= a_{ij} X_j + \sum_{h \neq j} a_{ih}e_h - T(X_j,a_{ij} X_j + \sum_{h \neq j} a_{ih}e_h) X_j\\
    &= a_{ij} X_j + \sum_{h \neq j} a_{ih}e_h - (a_{ij} + T(X_j,\sum_{h \neq j} a_{ih}e_h)) X_j\\
    &= a_{ij} X_j + \sum_{h \neq j} a_{ih}e_h - (a_{ij} + 0) X_j\\
    &= \sum_{h \neq j} a_{ih}e_h
\end{align*}
where in the second equality we substituted $X_i$, in the third equality we used regression equivariance and in the fourth equality we used the independence property. We obtain that $X_j$ is independent of $r_i^{(j)}$ for all $i \neq j$ as $X_j$ is independent of $\sum_{h \neq j} a_{ih}e_h$.\\
\\
\noindent 
\underline{2) Assume $X_j$ is not exogenous:}\\
If $X_j$ is endogenous, then there exists an exogenous variable $X_h = e_h$ such that $X_h$ has a directed path to $X_j$:
\begin{align*}
    r_h^{(j)} &= X_h - T(X_j,X_h)X_j\\
    &= e_h - T(X_j,X_h)\left( \sum_{k(t)<k(j)} a_{jt}e_t + e_j\right)\\
    &= (1 - T(X_j,X_h)a_{jh})e_h - T(X_j,X_h) \sum_{k(t)<k(j), t \neq h} a_{jt}e_t - T(X_j,X_h) e_j\\
    \text{and} \qquad &\qquad \\
    X_j &= \sum_{k(t)<k(j)} a_{jt}e_t + e_j = \sum_{k(t)<k(j),t \neq h} a_{jt}e_t + a_{jh} e_h + e_j
\end{align*}
If we now proof that $T(X_j,X_h)$ is nonzero, then the Darmois-Skitovitch theorem gives us that $r_h^{(j)}$ and $X_j$ are dependent as all the $e_k$ are independent and non-Gaussian. For this, we proceed as follows. First we have that
\begin{equation*}
    T(X_h,X_j) = T(e_h, \sum_{k(t)<k(j),t \neq h} a_{jt}e_t + a_{jh} e_h + e_j) = a_{jh}
\end{equation*}
using independence and regression equivariance. Second one has that
\begin{align*}
    \text{cov}(X_h,X_j) &= \text{cov}(e_h,\sum_{k(t)<k(j),t \neq h} a_{jt}e_t + a_{jh} e_h + e_j) = a_{jh} \text{var}(e_h)\\
    \implies a_{jh} &= \text{cov}(X_h,X_j)/\text{var}(e_h) \text{ as var}(e_h) > 0 
\end{align*}
Here $\text{cov}(X_h,X_j)$ cannot be zero under the correlation-faithfulness assumption as we have a directed path connecting them. Hence $T(X_h,X_j) = a_{jh}$ is nonzero. If we now use the third assumption in \eqref{eq:assumpslope} we have that $T(X_j,X_h)$ is also nonzero and we are done.\\
\\
For the Theil-Sen slope assumptions 1 and 2 from \eqref{eq:assumpslope} are immediately satisfied by using the stronger property of Fisher consistency. For the third assumption we proceed as follows. First we note that the Theil-Sen slope uses medians to estimate the regression slope between variables $Y$ and $X$. For these we have that the median of $\frac{Y-Y'}{X - X'}$ is nonzero if and only if the median of $\frac{X-X'}{Y - Y'}$ is nonzero. To see this: suppose without loss of generality that the median of the slopes $\frac{Y-Y'}{X - X'}$ is larger than zero. Then, as all
slopes larger than zero yield an inverse slope larger than zero, and likewise all
slopes smaller than zero yield an inverse slope smaller than zero, the median of $\frac{X-X'}{Y - Y'}$ is also larger than zero. Hence the median preserves the sign when swapping numerator and denominator and as we assume continuous variables, division by zero when swapping occurs with a negligible probability of zero. Hence $T(X,Y) \neq 0 \implies T(Y,X) \neq 0$ and we are done.

\end{proof}

\subsection{Proof of Lemma \ref{lemma:generalizedlemma2}}

\begin{proof}
$ $\newline
Assume that the mixing matrix $A$ in $X = A e$ has already been permuted to lower triangularity with ones on the diagonal and assume without loss of generality that $X_j = X_1 = e_1$. Since $X_1$ is exogenous, we have that the $a_{i1}$ for $i \neq 1$ are the slope coefficients when $X_i$ is regressed on $X_1$ using $T$:
\begin{align*}
    T(X_1,X_i) &= T(e_1, \sum_{t \leq i} a_{it}e_t)\\
    &= a_{i1} + T(e_1 , \sum_{t \leq i, t \neq 1} a_{it}e_t) \\
    &= a_{i1} + 0 = a_{i1}
\end{align*}
where we used regression equivariance in the second equality and independence in the third equality. Hence when we remove the effect of $X_1$ from $X_i$ by switching to the residuals $r_i^{(1)} = X_i - T(X_1,X_i) X_1 =  X_i - a_{i1} e_1$, the first column of $A$ becomes a zero vector. As $r^{(1)}$ is independent of $X_1$, we get for $r^{(1)}$ a new $(p-1) \times (p-1)$ dimensional lower triangular matrix $A^{(1)} = [A]_{2:p,2:p}$ with ones on the diagonal: $r^{(1)} = A^{(1)} e^{(1)}$  with $e^{(1)} = [e]_{2:p}$. Therefore a LiNGAM holds for the residual vector $r^{(1)}$.

Also, when switching to $r^{(1)}$, the corresponding matrix $A^{(1)}$ is the lower triangular submatrix formed by removing the first row and the first column of $A$. Hereby the causal order is not altered and hence switching to the residuals preserves the causal order.
\end{proof}

\subsection{Proof of Theorem \ref{thm:TSfishcons}}

\begin{proof}
$ $\newline
We have that: 
\begin{align*}
    &Y-Y' = \beta X + \varepsilon - \beta X' - \varepsilon' = \beta (X-X') + \varepsilon - \varepsilon', \quad \text{ with } \varepsilon \mydistr \varepsilon'\\
    &\implies \frac{Y-Y'}{X-X'}= \beta + \frac{\varepsilon - \varepsilon'}{X - X'}
\end{align*}
Hence the Theil-Sen slope is Fisher-consistent:
\begin{align*}
    &\iff \med \left( \frac{\varepsilon - \varepsilon'}{X - X'} \right) = 0 \qquad \qquad \qquad \qquad \\
    &\iff \mathbb{P}\left( \frac{\varepsilon-\varepsilon'}{X-X'} \leq 0\right) = 0.5\\
\end{align*}
Now for $\frac{\varepsilon - \varepsilon'}{X-X'}$ holds that $\varepsilon - \varepsilon'$ and $X-X'$ are symmetric about zero. Therefore $\frac{\varepsilon - \varepsilon'}{X-X'}$ is also symmetric about zero as numerator and denominator are independent and symmetric about zero, and thus we obtain that $\mathbb{P}\left( \frac{\varepsilon-\varepsilon'}{X-X'} \leq 0\right) = 0.5$.
\end{proof}

\subsection{Proof of Theorem \ref{thm:RMfishcons}}

\begin{proof}
    $ $\newline
    For $ X \mydistr X', \ \varepsilon \mydistr \varepsilon'$ independent, continuous random variables, we have that
    \begin{align*}
        Y - Y' &= \beta X + \varepsilon - \beta X' - \varepsilon' = \beta (X-X') + \varepsilon - \varepsilon'\\
        \implies \frac{Y - Y'}{X - X'} &= \beta + \frac{\varepsilon - \varepsilon'}{X - X'}
    \end{align*}
    Hence, for Fisher consistency of the repeated median, it is needed that: 
    \begin{align*}
        &\underset{X,\varepsilon}{\med} \left( \underset{X',\varepsilon'}{\med} \left( \frac{\varepsilon - \varepsilon'}{X-X'} \right) \right) = 0\\
        \iff &\mathbb{P}_{X,\varepsilon} \left( \underset{X',\varepsilon'}{\med} \left (\frac{\varepsilon - \varepsilon'}{X - X'} \right) \leq 0 \right) = 0.5\\
        \iff &\mathbb{P}_{X,\varepsilon} \left(x,e: \left[ \underset{X',\varepsilon'}{\med} \left (\frac{e - \varepsilon'}{x - X'} \right) \leq 0 \right] \right) = 0.5
    \end{align*}
    We compute:
    \begin{align*}
        &\underset{X',\varepsilon'}{\med} \left (\frac{e - \varepsilon'}{x - X'} \right) \leq 0
        \iff  \mathbb{P}_{X',\varepsilon'} \left( \frac{e - \varepsilon'}{x - X'} \leq 0 \right) \geq 0.5\\
        \iff & \mathbb{P} (e -\varepsilon' \leq 0 \ \cap \ x - X' \geq 0) + \mathbb{P} (e -\varepsilon' \geq 0 \ \cap \ x - X' \leq 0) \geq 0.5\\
        \iff &\mathbb{P} (e -\varepsilon' \leq 0) \cdot \mathbb{P}( x - X' \geq 0) + \mathbb{P} (e -\varepsilon' \geq 0 ) \cdot \mathbb{P}( x - X' \leq 0) \geq 0.5 \quad \text{[independence]}\\
        \iff &(1-F_{\varepsilon}(e)) \cdot F_X(x) + F_{\varepsilon}(e) \cdot (1-F_X(x)) \geq 0.5 \quad \text{[continuous r.v.]}
    \end{align*}
    This implies:
    \begin{align*}
        &\mathbb{P}_{X,\varepsilon} \left(x,e: \left[ \underset{X',\varepsilon'}{\med} \left (\frac{e - \varepsilon'}{x - X'} \right) \leq 0 \right] \right)\\
        = \ &\mathbb{P}_{X,\varepsilon} \Big(x,e: \ (1-F_{\varepsilon}(e)) \cdot F_X(x) + F_{\varepsilon}(e) \cdot (1-F_X(x)) \geq 0.5 \Big)\\
        = \ & \mathbb{P} \Big((1-F_{\varepsilon}(\varepsilon)) \cdot F_X(X) + F_{\varepsilon}(\varepsilon) \cdot (1-F_X(X)) \geq 0.5\Big)\\
        = \ & \mathbb{P} \Big( (1-U') U + U' (1-U) \geq 0.5\Big) \quad \text{with } U \coloneqq F_X(X), \ U' \coloneqq F_{\varepsilon}(\varepsilon) \text{ uniform and independent}\\
        \Big[ &(1-U') U + U' (1-U) \geq 0.5 \iff U \geq 0.5 \cap U' \leq 0.5 \text{ or } U \leq 0.5 \cap U' \geq 0.5 \Big] \\
        = \ & \int_{0.5}^1 \int_0^{0.5}dsdt +  \int_0^{0.5}\int_{0.5}^1dsdt = \frac{1}{2} \cdot \frac{1}{2} + \frac{1}{2} \cdot \frac{1}{2} = 0.5
    \end{align*}
    Hence the Repeated-Median is Fisher-consistent for continuous, random errors.
\end{proof}

\section{Additional simulation results}\label{app:simresults}

\begin{table}[h!]
\centering
\begin{tabular}{|r|rrrrr|rrrrr|}
  \hline
 sample size & 5 & 10 & 25 & 50 & 100 & 5 & 10 & 25 & 50 & 100\\ 
  \hline
  & &\multicolumn{3}{c}{$t_1$} & &  \multicolumn{5}{c|}{Pareto } \\
  \hline
  TSLiNGAM & \textbf{609} & 753 & \textbf{924} & \textbf{982} & \textbf{999} & \textbf{643} & \textbf{790} & \textbf{949} & \textbf{979} & \textbf{996}\\ 
DirectLiNGAM & 565 & 680 & 817 & 896 & 933 & 589 & 743 & 854 & 919 & 956\\
  EASE &  & 568 & 739 & 879 & 962 &  & 471 & 490 & 451 & 459\\  
  Repeated Median \& KBI & 582 & \textbf{758} & 905 & 974 & 996 & 609 & 740 & 904 & 956 & 991\\ 
  Theil-Sen \& dcorr & 571 & 744 & 920 & 977 & 994 & 608 & 778 & 943 & 977 & 995\\ 
  Repeated Median \& dcorr & 540 & 725 & 890 & 967 & 992 & 590 & 733 & 902 & 957 & 991\\ 
   \hline
  & &\multicolumn{3}{c}{$t_2$} & &  \multicolumn{5}{c|}{lognormal }  \\
  \hline
  TSLiNGAM & \textbf{568} & \textbf{639} & \textbf{795} & \textbf{908} & 960 & \textbf{624} & \textbf{761} & \textbf{925} & \textbf{971} & \textbf{996}\\ 
   DirectLiNGAM & 545 & 624 & 771 & 871 & 943 & 569 & 708 & 895 & 935 & 975\\ 
  EASE &  & 540 & 654 & 771 & 864 &  & 508 & 481 & 473 & 458\\ 
  Repeated Median \& KBI & 542 & 629 & 776 & 904 & 953 & 587 & 724 & 898 & 960 & 988\\ 
  Theil-Sen \& dcorr & 547 & 616 & 772 & 892 & \textbf{962} & 540 & 745 & 912 & 965 & 994\\ 
  Repeated Median \& dcorr & 519 & 625 & 744 & 885 & 954 & 545 & 700 & 882 & 951 & 984\\ 
  \hline
  & &\multicolumn{3}{c}{$t_5$} & &  \multicolumn{5}{c|}{exponential } \\ 
  \hline
  TSLiNGAM & \textbf{538} & 522 & \textbf{618} & \textbf{698} & 789 & \textbf{604} & \textbf{703} & 884 & \textbf{961} & 974\\ 
DirectLiNGAM & 517 & 513 & 616 & 692 & \textbf{796} & 572 & 676 & 861 & 926 & \textbf{980}\\ 
  EASE &  & \textbf{535} & 557 & 610 & 696 &  & 474 & 412 & 406 & 379\\ 
  Repeated Median \& KBI & 522 & 527 & 617 & 677 & \textbf{796} & 555 & 674 & 849 & 935 & 973\\ 
  Theil-Sen \& dcorr & 533 & 518 & 596 & 655 & 758 & 567 & 696 & \textbf{888} & 959 & \textbf{980}\\ 
  Repeated Median \& dcorr & 505 & 525 & 580 & 644 & 718 & 547 & 670 & 860 & 938 & 972\\ 
   \hline
\end{tabular}
\caption{Number of correct causal orders out of 1000 trials for p = 2.}
\label{tab:simulationp2}
\end{table}

Table \ref{tab:simulationp2} and \ref{tab:simulationp5} present the simulation results for $p=2$ and $p = 5$ variables respectively. It is clear that TSLiNGAM performs best overall. In particular, it outperforms DirectLiNGAM on heavy-tailed and skewed distributions, and the outperformance is more pronounced as the tails gets heavier. For lighter tails, the performance becomes similar to DirectLiNGAM. Using the repeated median works well, but provides no improvement over TSLiNGAM. The dcorr independence measure performs somewhat worse than the kernel-based independence measure, but becomes competitive at larger sample sizes. EASE is not doing very well, and needs larger sample sizes to become competitive.

\begin{table}[h!]
\centering
\begin{tabular}{|r|rrrr|rrrr|}
  \hline
sample size & 30 & 50 & 100 & 200 & 30 & 50 & 100 & 200 \\ 
  \hline
  & \multicolumn{4}{c|}{$t_1$} &   \multicolumn{4}{c|}{Pareto } \\
  \hline
  TSLiNGAM & \textbf{709} & \textbf{893} & \textbf{947} & \textbf{992} & \textbf{742} & \textbf{887} & \textbf{979} & \textbf{999} \\ 
DirectLiNGAM & 504 & 620 & 740 & 794 & 589 & 713 & 796 & 926 \\ 
  EASE & 183 & 399 & 630 & 826 & 88 & 119 & 146 & 206 \\ 
  Repeated Median \& KBI & 663 & 857 & 943 & 991 & 645 & 827 & 943 & 996 \\ 
  Theil-Sen \& dcorr & 567 & 769 & 895 & 960 & 677 & 815 & 959 & 995 \\ 
  Repeated Median \& dcorr & 528 & 738 & 872 & 951 & 557 & 743 & 919 & 986 \\ 
  \hline
  & \multicolumn{4}{c|}{$t_2$} &   \multicolumn{4}{c|}{lognormal } \\
  \hline
  TSLiNGAM & \textbf{405} & \textbf{551} & \textbf{747} & 853 & \textbf{688} & \textbf{847} & \textbf{954} & \textbf{990} \\ 
  DirectLiNGAM & 390 & 513 & 709 & 809 & 558 & 748 & 873 & 947 \\ 
  EASE & 128 & 171 & 297 & 495 & 79 & 84 & 112 & 169 \\ 
  Repeated Median \& KBI & 366 & 520 & 743 & 850 & 591 & 772 & 930 & 982 \\ 
  Theil-Sen \& dcorr & 283 & 435 & 688 & \textbf{892} & 600 & 779 & 929 & 986 \\ 
  Repeated Median \& dcorr & 264 & 396 & 652 & 859 & 487 & 679 & 885 & 964 \\ 
  \hline
  & \multicolumn{4}{c|}{$t_5$} &   \multicolumn{4}{c|}{exponential } \\
  \hline
  TSLiNGAM & 157 & 220 & 327 & 407 & \textbf{514} & \textbf{716} & 875 & 962 \\ 
  DirectLiNGAM & 158 & \textbf{225} & \textbf{333} & \textbf{429} & 502 & 679 & 846 & 950 \\ 
  EASE & 82 & 77 & 122 & 161 & 58 & 60 & 62 & 88 \\ 
  Repeated Median \& KBI & \textbf{160} & 217 & 322 & 394 & 458 & 647 & 852 & 940 \\ 
  Theil-Sen \& dcorr & 71 & 102 & 185 & 267 & 466 & 691 & \textbf{879} & \textbf{970} \\ 
  Repeated Median \& dcorr & 72 & 91 & 151 & 218 & 375 & 592 & 818 & 936 \\ 
   \hline
\end{tabular}
\caption{Number of correct causal orders out of 1000 trials for p = 5.}
\label{tab:simulationp5}
\end{table}

The computation times for the simulation with $p=2$ and $p=5$ variables with $t_5$ distributions are given in Tables \ref{tab:comptimep2} and \ref{tab:comptimep5} respectively. It is clear that TSLiNGAM and DirectLiNGAM have similar computational costs. Using dcorr as independence measure decreases the computational costs by a factor of roughly 3 for $p = 2$ and a factor of 10 for $p=5$. EASE is again by far the fastest method.

\begin{table}[h!]
\centering
\begin{tabular}{|r|rrrrr|}
  \hline
 sample size & 5 & 10 & 25 & 50 & 100 \\ 
  \hline
  TSLiNGAM & 3.42 & 4.09 & 4.99 & 5.84 & 7.32 \\ 
DirectLiNGAM & 3.55 & 4.13 & 4.98 & 5.70 & 6.71 \\ 
  EASE &  & 0.23 & 0.21 & 0.22 & 0.25 \\ 
  Repeated Median \& KBI & 3.47 & 4.19 & 5.16 & 6.25 & 8.02 \\ 
  Theil-Sen \& dcorr & 1.11 & 1.07 & 1.13 & 1.21 & 1.26 \\ 
  Repeated Median \& dcorr & 1.05 & 1.03 & 1.23 & 1.37 & 1.73 \\ 
   \hline
\end{tabular}
\caption{Computational time in seconds for 1000 runs for $p=2$ for a Student-$t$ distribution with 5 degrees of freedom.}
\label{tab:comptimep2}
\end{table}

\begin{table}[h!]
\centering
\begin{tabular}{|r|rrrr|}
  \hline
 sample size & 30 & 50 & 100 & 200 \\ 
  \hline
  TSLiNGAM & 1.60 & 1.80 & 2.19 & 2.88 \\ 
DirectLiNGAM & 1.57 & 1.78 & 2.12 & 2.81 \\ 
  EASE & 0.01 & 0.01 & 0.02 & 0.01 \\ 
  Repeated Median \& KBI & 1.66 & 1.93 & 2.40 & 3.18 \\ 
  Theil-Sen \& dcorr & 0.10 & 0.13 & 0.21 & 0.35 \\ 
  Repeated Median \& dcorr & 0.13 & 0.20 & 0.36 & 0.55 \\ 
   \hline
\end{tabular}
\caption{Computational time in minutes for 1000 runs for $p=5$ for a Student-$t$ distribution with 5 degrees of freedom.}
\label{tab:comptimep5}
\end{table}

\section{Additional figures}\label{app:additionalfigures}
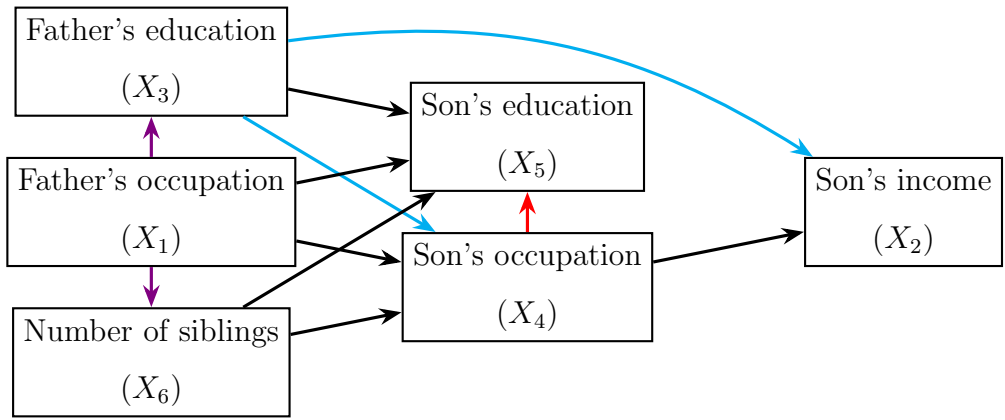
\begin{figure}[h!]
    \centering
    \begin{tikzpicture}
        \begin{scope}[every node/.style={circle,thick,draw}]
            \node[shape=rectangle,draw=black,align=center] (3) at (0,0) {Father's education\\
            $(X_3)$};
            \node[shape=rectangle,draw=black,align=center] (1) at (0,-2) {Father's occupation\\
            $(X_1)$};
            \node[shape=rectangle,draw=black,align=center] (6) at (0,-4) {Number of siblings\\
            $(X_6)$};
            \node[shape=rectangle,draw=black,align=center] (5) at (5,-1) {Son's education\\
            $(X_5)$};
            \node[shape=rectangle,draw=black,align=center] (4) at (5,-3) {Son's occupation\\
            $(X_4)$};
            \node[shape=rectangle,draw=black,align=center] (2) at (10,-2) {Son's income\\
            $(X_2)$};
        \end{scope}

        \begin{scope}[>={Stealth[black]},
              every edge/.style={draw,very thick,}]

            \path [-{Stealth[color=violet]},draw=violet] (1) edge node {} (3);
            \path [-{Stealth[color=violet]},draw=violet] (1) edge node {} (6);
            \path [-{Stealth[color=cyan]},draw=cyan] (3) edge[bend left=20] node {} (2);
            \path [-{Stealth[color=cyan]},draw=cyan] (3) edge node {} (4);
            \path [->] (3) edge node {} (5);
            \path [->] (1) edge node {} (5);
            \path [->] (1) edge node {} (4);
            \path [->] (6) edge node {} (5);
            \path [->] (6) edge node {} (4);
            \path [->] (4) edge node {} (2);
            \path [-{Stealth[color=red]},draw=red] (4) edge node {} (5);

        \end{scope}
    \end{tikzpicture}
    \captionsetup{width=0.85\textwidth}
    \caption{Causal graph found by Theil-Sen and distance correlation with correct arrows in black, wrongly directed arrows in red, unverifiable arrows in purple and redundant arrows in blue.}
    \label{fig:DAGTSdcorsoc}
\end{figure}

\end{document}